\documentclass[journal]{IEEEtran}
\usepackage{upgreek}
\usepackage{graphicx}
\usepackage{subfigure}
\usepackage{amsmath,bm}
\usepackage{amssymb} %for leqslant
\usepackage{mathrsfs} %花体

\usepackage{enumerate}
\usepackage[square, comma, sort&compress, numbers]{natbib}
\usepackage{epstopdf}
\usepackage{color}
\ifCLASSINFOpdf
\else
\fi

\begin{document}

\title{%Mobile Energy Transfer \\ via Self-Aligned Radiative Resonances
%Self-Aligned Resonant Beam Charging for \\ A Mobile IoT Device
Charging A Smartphone Over the Air: The Resonant Beam Charging Method
}

% Place the author information here.  Please hand-code the contact
% information and notecalls; do *not* use \footnote commands.  Let the
% author contact information appear immediately below the author names
% as shown.  We would also prefer that you don't change the type-size
% settings shown here.

\author{\normalsize{Qingwen~Liu\#*,~\IEEEmembership{\normalsize Senior Member,~IEEE\normalsize},
Mingliang~Xiong\#, Mingqing~Liu\#,
Qingwei~Jiang, Wen~Fang, Yunfeng~Bai }
%\thanks{Qingwen Liu, Mingliang Xiong, Mingqing Liu contribut equally to this work. The corresponding author is Qingwen Liu.}
\thanks{
\# Qingwen Liu, Mingliang Xiong, and Mingqing Liu contributed equally to this work.

* The corresponding author: Qingwen Liu.

Qingwen Liu, Mingliang Xiong, Mingqing Liu, Qingwei Jiang, Wen Fang, and Yunfeng Bai are with College of Electronics and Information Engineering, Tongji University, Shanghai, People's Republic of China (e-mail: qliu@tongji.edu.cn, xiongml@tongji.edu.cn, clare@tongji.edu.cn, jiangqw@tongji.edu.cn, wen.fang@tongji.edu.cn, baiyf@tongji.edu.cn)

This work was also supported by the National Natural Science Foundation of China under Grant 62071334 and Grant 61771344, and the National Key Research and Development Project under Grant 2020YFB2103902.}
}

	\maketitle

	\begin{abstract}
%Charging a smartphone remains the daily anxiety because it is so difficult to charge a smartphone in mobile operation.

Wireless charging for mobile Internet of Things (IoT) devices such as smartphones is extremely difficult. To reduce energy dissipation during wireless transmission in mobile scenarios,  laser or narrow radio beams with sophisticated tracking control are typically required. However, reaching the necessary tracking accuracy and reliability is really difficult. In this paper, inspired by the features of optical resonators and retroreflectors, we develop an experiment on a self-aligned resonant beam charging system for long-distance mobile power transfer. It exploits light resonances inside a double-retroreflector-based spatially separated laser resonator~(SSLR), which eliminates the requirement for any kind of tracking control. Focal telecentric cat's eye retroreflectors are employed here. The SSLR was investigated by both theoretical calculation and experiment. We also well assembled the transmitter and the receiver and demonstrated its application in mobile smartphone charging. The results show that above $5\mbox{-W}$ optical power~(also obtained more than $0.6\mbox{-W}$ electrical power) transferring with negligible diffraction loss to a few-centimeter-size receiver is realized while the receiver moves arbitrarily within $2\mbox{-m}$ vertical distance and $6^\circ$ field of view from the transmitter. The maximum horizontal moving range is up to $\pm 18~\mbox{cm}$. This wireless charging system empowers a smartphone in mobile operation with unlimited battery life without the need for a cable.

%Wireless power transfer is leading a revolution in energy distribution. However, it faces a fundamental challenge of usually requiring a sophisticated tracking mechanism to realize long-distance and high-efficiency transfer in mobile scenarios. By exploiting radiative resonances, we present theoretically and verify experimentally that a resonant light beam inside the double-retroreflector cavity acts as a self-aligned channel for mobile energy transfer. Without tracking control, we transfer above 5-watt optical power with nearly 100$\boldsymbol{\boldsymbol{\%}}$ efficiency to a few-centimeter-size receiver, which is moving arbitrarily in the range of 2-meter distance and 6-degree field of view from the transmitter. We further demonstrate that this energy supply empowers a smartphone in mobile operation with unlimited battery life. The mechanism of self-aligned radiative resonances may apply in a general system of physical resonances, which can facilitate scientific innovations and imaginative applications.
		
	\end{abstract}

\begin{IEEEkeywords}
Mobile wireless power transfer, laser charging, distributed laser charging, spatially separated laser resonator.
%, Internet of Things.
\end{IEEEkeywords}

	% In setting up this template for *Science* papers, we've used both
	% the \section* command and the \paragraph* command for topical
	% divisions.  Which you use will of course depend on the type of paper
	% you're writing.  Review Articles tend to have displayed headings, for
	% which \section* is more appropriate; Research Articles, when they have
	% formal topical divisions at all, tend to signal them with bold text
	% that runs into the paragraph, for which \paragraph* is the right
	% choice. Either way, use the asterisk (*) modifier, as shown, to
	
	\section{Introduction}
		\begin{figure}[t]
		\centering
		\includegraphics[width=3.4in]{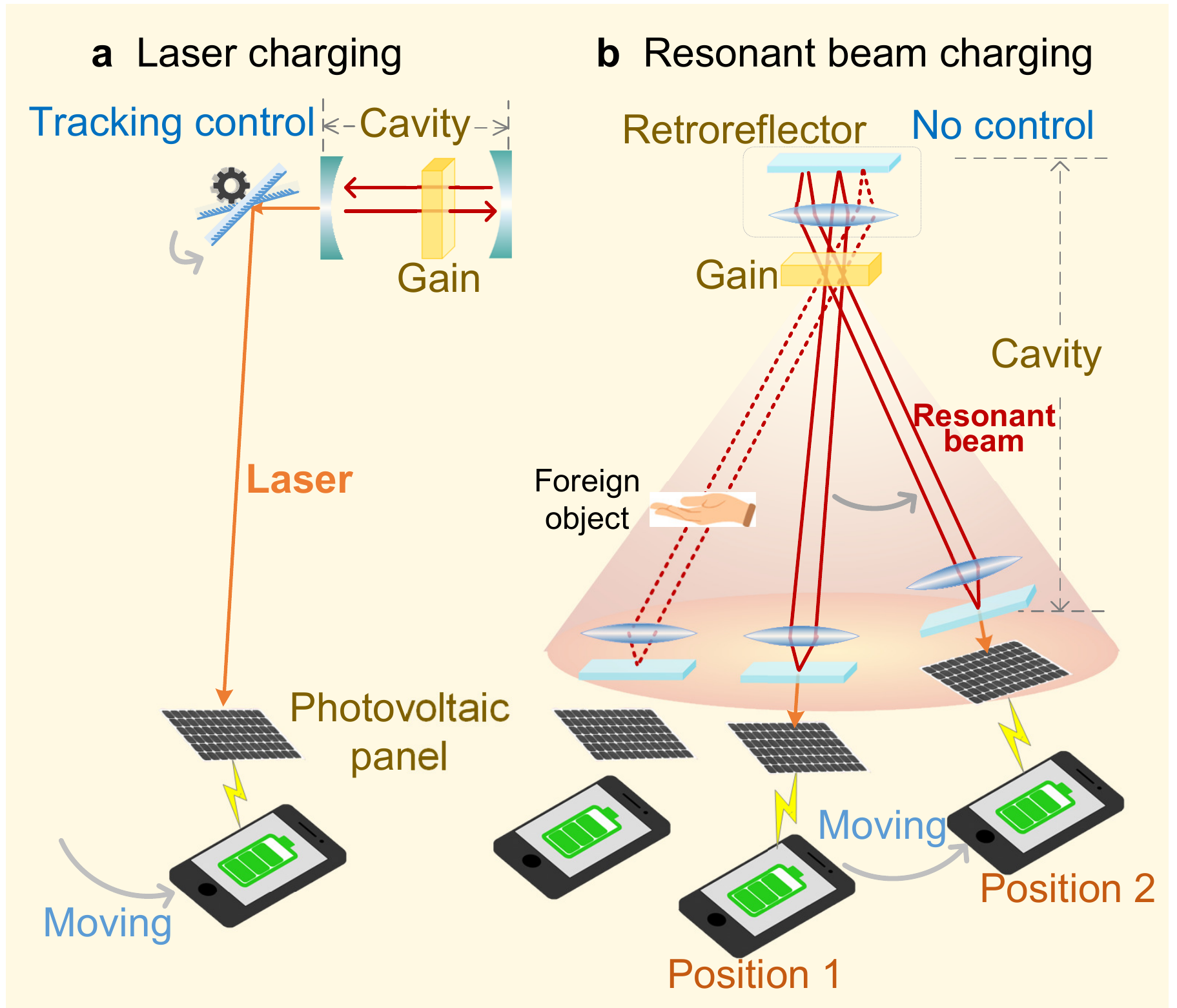}
		\caption{Comparison of laser charging and self-aligned resonant beam charging.
			\textbf{a}, Laser charging. It needs tracking a moving electronic device to maintain high efficiency.
			\textbf{b}, Self-aligned resonant beam charging. The beam is self-aligned due to the double-retroreflector design for automatically and efficiently charging a moving electronic device without tracking control.}
		\label{f:comparison}
	\end{figure}
	Nikola Tesla's pioneering contributions to wireless power transfer (WPT) have inspired researchers to convert scientific discovery into engineering practice for more than a century~\cite{Nikola1914APPARATUS}. In the past decade, the tremendous emergence of mobile IoT devices (e.g., smartphones, smartwatches, robots, electric vehicles, etc.) has led to the rapid growth of WPT technologies, which could eliminate their troublesome cables and  batteries~\cite{7096295}. Non-radiative near-field WPT realized by coils is suitable for high-efficiency transfer; however, it faces the challenge that the effective transmission distance is limited within several times of transceivers' sizes, even if the operation is in the coupled magnetic resonating regime~\cite{ GeoffroyNature,Kurs2007Wireless,2017Robust}. Radiative far-field WPT (e.g., radio wave, lightwave, and laser; see Fig.~\ref{f:comparison}a) is natural for long-range transfer, but it has the difficulty that  high-efficiency transfer requires a sophisticated mechanism for receiver positioning and tracking, since
	reaching necessary tracking accuracy and reliability is a severe challenge or even insurmountable obstacle~\cite{GaASPV,LaserSLIPT2,SWIPT1,RFCharging}.
	
	Intuitively, if two  objects exchange energy in a resonant regime, the energy transfer is expected to be  high-efficiency since the  resonant field is formed by the superposition of in-phase waves (other phases are canceled by superposition)~\cite{coupleResonance}. The mechanism of resonances may apply in a general physical system (e.g., acoustic, electromagnetic, astronomic, nuclear, etc.). Thus, exploiting far-field radiative resonance, simultaneously long-range and high-efficiency WPT is expected to be realized. Here, we focus on one particular physical embodiment: optical resonances~\cite{JMLiuOptRes}. Optical resonances are suitable for common applications because nearly no interference involves with most electronic devices and interactions with environmental objects are limited due to their spatially-concentrated field distribution.

	Relying on optical resonances of a separated two-mirror resonator, Wang \emph{et al.} demonstrated a WPT system up to $2\mbox{-m}$ transmission distance~\cite{WWang2018}. While the mirrors in~\cite{WWang2018} need to be aligned accurately and manually. To achieve mobility without any  alignment or tracking control, recent papers presented detailed theoretical analysis and numerical simulation on the feasibility of using a double-retroreflector-based spatially separated laser resonator~(SSLR) for beam self-alignment/self-tracking; see Fig.~\ref{f:comparison}b~\cite{xiongretro,liucateye,liuretro}. As depicted in Fig.~\ref{f:comparison}b, both the transmitter and the receiver jointly constitute the SSLR. The standing-wave light beam in the SSLR is termed as the \emph{resonant beam}. Both corner cube retroreflectors and telecentric cat's eye retroreflectors~(TCRs) were investigated in recent papers. Spacifically in \cite{xiongretro}, a stable SSLR based on focal telecentric cat's eye retroreflectors~(FTCRs) was proposed to provide a very low diffraction loss over a large range of cavity length. As in theoretical analysis, long-range and mobile energy transfer implemented in this way can be high-efficiency with low interference into environmental objects. Nevertheless, many detailed design information and parameters should be determined for realizing such a sophisticated structure in practice. Also, many practical issues such as thermal distortion, inhomogeneous intensity, and beam collection at the arbitrary direction are expected to be overcome by the experiment.
	
	In this paper, we analyze theoretically and demonstrate experimentally a complete mobile resonant beam charging system, including the whole progress: pump current input, resonant beam generation and transmission, beam collection, photovoltaic conversion, and charging a smartphone. All the aforementioned practical issues are addressed.  By exploring the quantified analysis of
electromagnetic signal, we make the following contributions:

\begin{itemize}
\item[1)] To the best of our knowledge, this is the first time demonstrating experimentally a complete mobile WPT system with up to 2-m distance, multi-Watt  optical power, centimeter-level receiving area, and $5$-V charging voltage. The safety of foreign object invasion is also verified.

\item[2)] Theoretical analysis and experimental measurement are well-matched in this work. Parameter estimation is also conducted to determine several unmeasured parameters in this system. We use an analytical model for calculating
the over-the-air transmission efficiency and the deliverable power of the proposed wireless power transfer scheme to demonstrate its low diffraction loss. The proposed analytical model in this paper needs fewer assumptions compared to the previously presented models.
\end{itemize}

The remainder of this paper is organized as follows. In Section II, we present the architecture of the proposed system and the analytical model for calculating the over-the-air transmission efficiency and the charging power. In Section III, we conduct both experiments and numerical calculations to prove the self-alignment characteristic and quantitatively analyze the charging/moving performance of the system. Afterwards, we discuss several related aspects, such as improving mobility, efficiency, safety, and extensibility in Section IV. Finally, we make a conclusion in Section V.

	\section{Model and Analysis}
	
	%For example, the optical resonant beam system using the curve-mirror cavity has exhibited high-efficiency WPT for a fixed receiver\cite{WWang2018}.
	%The resonant beam for atmospheric pollution detection has been demonstrated, which was created between two retroreflectors for convenient optical alignment\cite{Linford1,Linford2}. Here, we explore the resonant beam in the double-retroreflector cavity, which enables WPT for a moving receiver.
	\begin{figure}
		\centering
		\includegraphics[width=3.4in]{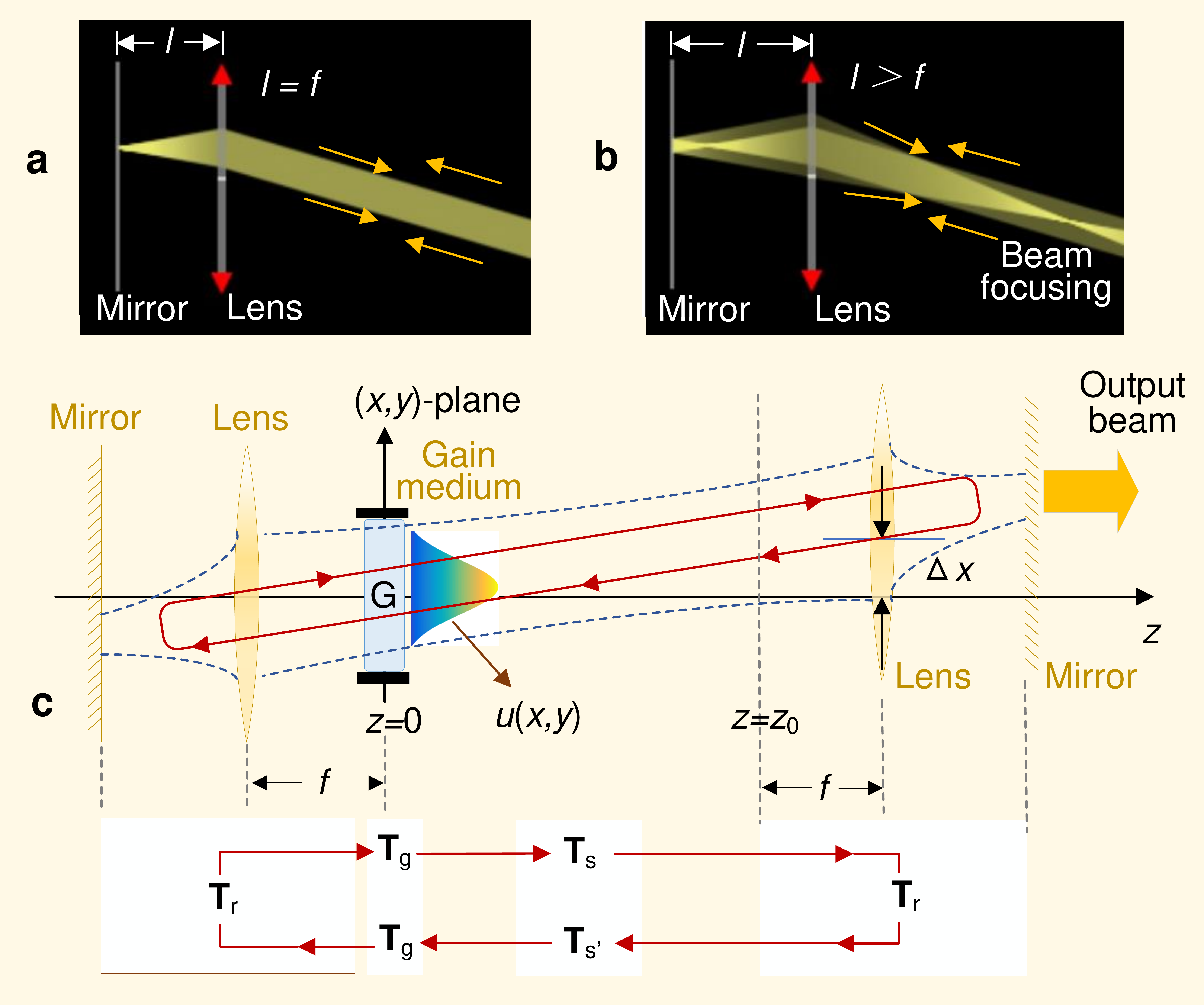}
		\caption{
				Theoretical model and analysis. \textbf{a}, Conventional telecentric cat's eye retroreflector~(TCR). The interval between mirror and convex lens $l$ equals the convex lens' focal length $f$ ($l=f$).
			\textbf{b}, Focal telecentric cat's eye retroreflector~(FTCR). The $l>f$  setting enables beam focusing for cavity stability. \textbf{c}, Resonant beam system based on FTCRs. The self-reproducing mode theory model is illustrated by the transition matrices of field transfer for one round trip (red arrow line). $\Delta x$ is the relative movement along x-axis of the retroreflector at the receiver.  	}
		\label{f:transfer}
	\end{figure}
%\textbf{d}, Over-the-air transmission efficiency $\eta_{\rm t}$ vs. receiver movement along x-axis $\Delta x$ from $(x_0=0,y_0=0,z_0=2~\mbox{m})$. \textbf{e}, $\eta_{\rm t}$ vs. receiver movement along z-axis $\Delta z$ from  $(x_0=0,y_0=0,z_0=0)$.
	To describe how the resonant beam in the double-retroreflector cavity leads to a self-alignment ability, we adopt an analysis method based on the self-reproducing mode theory and numerical integration~\cite{hodgson2005laser,FoxLi,FFT,Fresnel}.
	
	\subsection{Focal telecentric cat's eye retroreflector}
	
	 Figure~\ref{f:transfer}a shows that a conventional TCR consists of a plane mirror and a convex lens, which are parallel to each other, where the interval between them equals the focal length of the convex lens. It can reflect the incident beam from an arbitrary direction back parallel to the original direction~\cite{cateye1}. In this wireless charging system, we adopt FTCRs, as depicted in Fig.~\ref{f:transfer}b, to ensure the stability of the SSLR~\cite{xiongretro}. The distance between the mirror and the convex lens is slightly larger than the focal length of the convex lens, which exhibits a beam focusing ability.
	
	 Using the matrix optics method, the process that a ray passes through an optical component can be described as the ray vector multiplied by a matrix under the paraxial approximation; that is~\cite{matrix}
	\begin{equation}
	\left[\begin{array}{l}
	x_{\rm o} \\
	\theta_{\rm o}
	\end{array}\right]=\mathbf{M_0}\left[\begin{array}{l}
	x_{\rm i} \\
	\theta_{\rm i}
	%r_{\mathrm{o}} \\
	%\alpha_{\mathrm{o}}
	%\end{array}\right]=\mathbf{M}\left[\begin{array}{l}
	%r_{\mathrm{i}} \\
	%\alpha_{\mathrm{i}}
	\end{array}\right],
	\end{equation}
	where $x_{\rm i}$ and $x_{\rm o}$ are the transverse displacements of the input ray and output ray from the optical axis, respectively, while $\theta_{\rm i}$ and $\theta_{\rm o}$ are the slopes of the input ray and output ray, respectively.
	%while $\alpha_{\mathrm{i}}$ and $\alpha_{\mathrm{o}}$
	
	Moreover, if the ray passes through a series of optical components, the overall ray-transfer matrix is the production of individual matrices of these components with the opposite order~\cite{matrix2,matrix3}. Thus, the ray-transfer matrix of the FTCR describes the  process: A ray inputs at the incident plane, passes through all optical elements in a sequence, and finally outputs from the incident plane; it is expressed as~\cite{xiongretro}
	\begin{equation}\label{matrix1}
	\begin{aligned}
	\mathbf{M} =&\left[\begin{array}{ll}
	1 & f \\
	0 & 1
	\end{array}\right]\left[\begin{array}{cc}
	1 & 0 \\
	-1 / f & 1
	\end{array}\right]\left[\begin{array}{ll}
	1 & l \\
	0 & 1
	\end{array}\right]\left[\begin{array}{ll}
	1 & 0 \\
	0 & 1
	\end{array}\right] \\
	&\left[\begin{array}{ll}
	1 & l \\
	0 & 1
	\end{array}\right]\left[\begin{array}{cc}
	1 & 0 \\
	-1 / f & 1
	\end{array}\right]\left[\begin{array}{ll}
	1 & f \\
	0 & 1
	\end{array}\right] \\
	=&\left[\begin{array}{cc}
	1 & 0 \\
	-1 / f_{\mathrm{RR}} & 1
	\end{array}\right]\left[\begin{array}{cc}
	-1 & 0 \\
	0 & -1
	\end{array}\right],
	\end{aligned}
	\end{equation}
	where
	\begin{equation}
	f_{\mathrm{RR}}=1 \Big{/}\left(\frac{2 l}{f^{2}}-\frac{2}{f}\right).
	\end{equation}
	%Thus, as the left and right part in $\mathbf{M}$ indicate, the proposed retroreflector integrates the functions of a lens and a mirror, respectively.
	$f$ is the lens's focal length, and $l$ is the interval between the plane mirror and the lens in the retroreflector.
	If $l = f$ (see Fig.~2b), the equation \eqref{matrix1}  becomes
	\begin{equation}
	\mathbf{M_c} =  \left[\begin{array}{cc}
	-1 & 0 \\
	0 & -1
	\end{array}\right].
	\end{equation}
	Thus, $x_{\rm o}=-x_{\rm i}$ and $\theta_{\rm o}=-\theta_{\rm i}$, i.e., the input ray passes through the reflector will return to the reverse direction. If $l>f$ (see Fig.~2b), the equation \eqref{matrix1} becomes
	\begin{equation}
	\mathbf{M_r} =  \left[\begin{array}{cc}
	-1 & 0 \\
	1/f_{\mathrm{RR}} & -1
	\end{array}\right].
	\end{equation}
	Thus, $x_{\rm o}=-x_{\rm i}$ and $\theta_{\rm o}=-\theta_{\rm i}+1/f_{\mathrm{RR}}$, i.e., the retroreflector exhibits the capability of beam focusing.
	
\begin{figure}[t]
		\centering
		\includegraphics[width=3.5in]{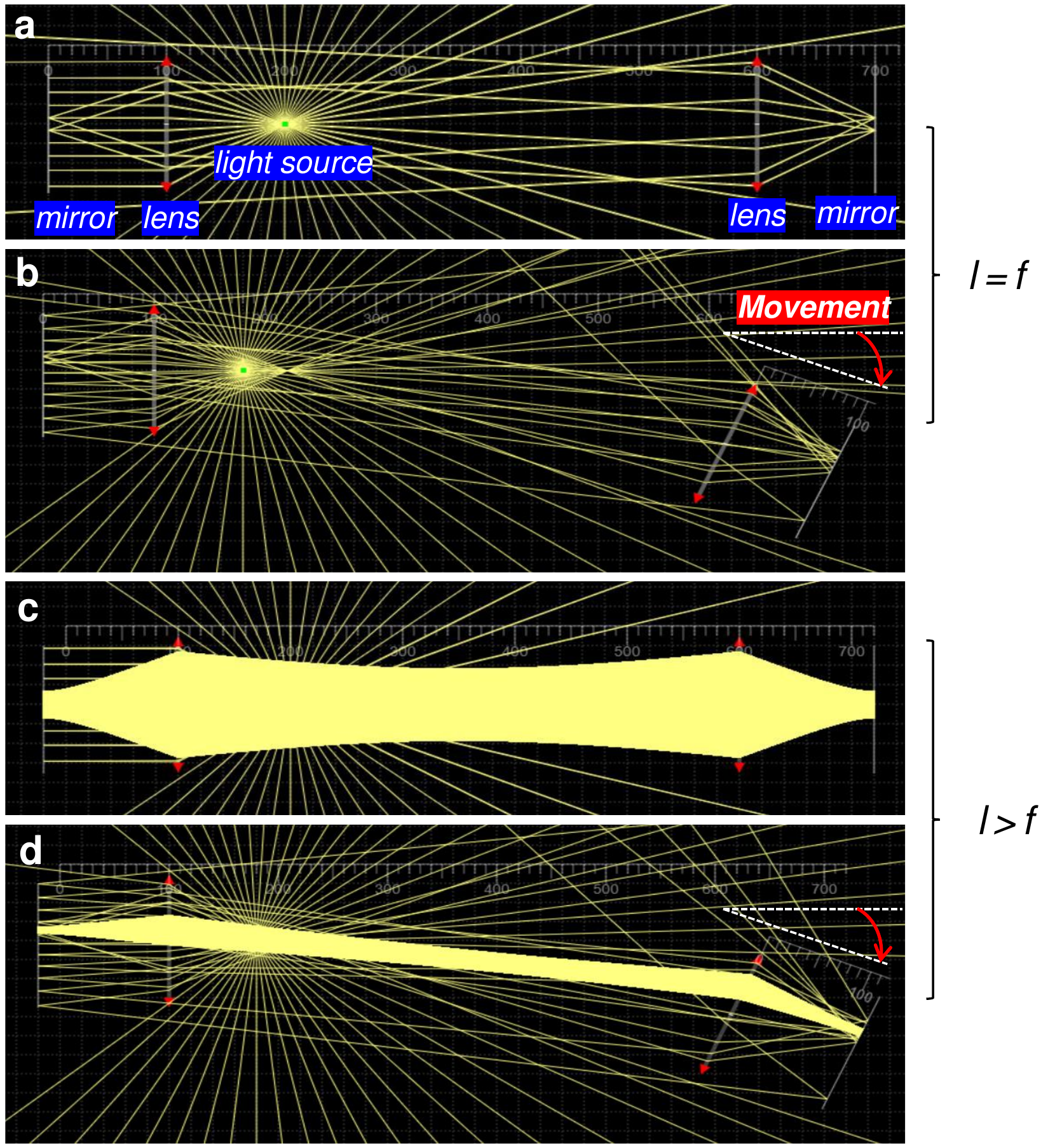}
		\caption{Ray-Optics simulation for ray tracing compares the conventional TCR-based cavity and the FTCR-based cavity. \textbf{a}, conventional TCR-based cavity. A light source is set at the lens's focal point of a retroreflector and when the mirror-lens interval $l$ equals the focal length of the lens $f$, i.e., $l = f$, the ray tracing can not form a stable oscillation in the cavity (the program stops running only when all the rays fly out of the scope). \textbf{b}, Conventional TCR-based cavity after movement. The stable oscillation cannot be formed. \textbf{c}, FTCR-based cavity with $l > f$. The rays in the cavity can circulate back and forth in the cavity without overflow and form a stable beam. \textbf{d}, FTCR-based cavity with $l > f$ after movement. The stable oscillation beam can still be formed.}
		\label{f:raytrace}
	\end{figure}

	We use a ray-tracing optics simulator Ray-Optics (https://github.com/ricktu288/ray-optics) to illustrate that the FTCR-based SSLR is more stable than the TCR-based cavity, as shown in Fig.~\ref{f:raytrace}. We can see that the SSLR based on conventional TCRs can only capture rays in a specific direction, while the SSLR based on FTCRs can capture  many rays although they have different paths.
	
	Figure~\ref{f:transfer}c %and Fig.~\ref{f:comparison}B
	shows the SSLR-based resonant beam transfer system with a transmitter and a receiver. The transmitter includes an FTCR and a gain medium. The receiver consists of the other FTCR. The collaboration between these two retroreflectors leads to the self-alignment feature of the cavity, where the resonant beam can be established between these two retroreflectors, even when the relative position of the transmitter and the receiver changes. In addition, the stability of the cavity is enhanced due to the beam-focusing ability of the FTCR. Based on the laser principle, the gain medium acts as a power amplifier fed by input power and stimulates a resonant beam. If the retroreflector at the receiver is partially reflective (reflectivity is $R$), the resonant beam can partly pass through to form an output beam, i.e., laser. Then, the output beam is converted into electricity by a photovoltaic~(PV) panel.
	
	%Based on laser principle, power is fed into the gain medium with the gain to stimulate the intracavity resonant beam power. {\color{black}If the round trip optical gain can compensate for the transfer losses within the cavity as $g_{0}^2R_tR_rT^2(1-\eta_{\rm t}) \geq 1$, the oscillation can be built up~\cite{Linford2}. $g_0$ is the small signal laser amplifier gain, $R_t$ and $R_r$ are reflectivities of the retroreflectors at the transmitter and the receiver, $T^2$ is the cavity transmission, and $1-\eta_{\rm t}$ represents the path loss over the air, where $\eta_{\rm t}$ is the transmission efficiency of the intracavity resonant beam. Only when the above oscillation threshold condition is satisfied, the stable resonant beam can be established as an energy transfer channel, where the output beam, i.e., laser beam, at the receiver is converted to electricity by a photovoltaic panel.}
	
	%\section{Theoretical Analysis}
	\subsection{Power transfer from the transmitter to the receiver}
	In such a resonant beam charging system, the energy flow can be divided into four stages: i) the input power $P_{\rm in}$ at the transmitter is absorbed\ by the gain medium to stimulate a resonant beam; ii) the resonant beam transfers over the air to the receiver;  iii) the receiver's retroreflector outputs parts of the resonant beam; and iv) the output beam is converted into electrical power $P_{\rm out}$ for smartphone charging. Thus, the relationship between $P_{\rm in}$ and $P_{\rm out}$ can be depicted as~\cite{WWang2018}
	\begin{equation}
	\begin{aligned}
	P_{\rm out} &= \eta_{\rm pv}\eta_{\rm e}\eta_{\rm g}(P_{\rm in}-P_{\rm th}),\\
	\eta_{\rm e}&= F(\eta_{\rm t}),
	%P_{\rm out} = \eta_{extr}(\eta_{exci}P_{\rm in}-C),
	%P_{\rm out} = \eta_{\rm pv}\eta_{ex}\eta_{\rm g}(P_{\rm in}-P_{\rm th}),
	\label{e:power}
	\end{aligned}
	\end{equation}
	where $\eta_{\rm pv}$ is the photovoltaic efficiency, $\eta_{\rm e}$ is a function of $\eta_{\rm t}$ indicated by $F(\cdot)$, $\eta_{\rm t}$ is the over-the-air transmission efficiency, $\eta_{\rm g}$ is the gain-medium stimulation efficiency, and $P_{\rm th}$ is the input power threshold of enabling resonance. $\eta_{\rm pv}$ and $\eta_{\rm g}$ rely on hardware, while $\eta_{\rm t}$ depends on the relative position between the transmitter and the receiver.

	Based on the laser principle, power is fed into the gain medium to stimulate the intra-cavity resonant beam power~\cite{Koechner2013Solid,hodgson2005laser,siegmanlaser}. If the gain can compensate the transfer losses for one round trip within the cavity, i.e., $\exp{(2g_{0}\ell_{\rm g}) }\geq \delta$, the resonance can be built up~\cite{Koechner2013Solid}. $g_0$ is the small-signal gain coefficient, $\ell_{\rm g}$ is the path length of the beam passing through the gain medium, and $\delta$ represents the total round trip path loss factor including the over-the-air transmission efficiency $\eta_{\rm t}$ of the intra-cavity resonant beam, the loss factor of passing inside the gain medium, and the reflectivity of the retroreflector at the receiver. Only when the above resonance threshold condition is satisfied can the intra-cavity resonant beam be established as an energy transfer channel. The output laser beam at the receiver is converted into electricity by a PV panel.
	$P_{\rm th}$ in equation (1) is the threshold power for initializing the resonance, i.e., enabling $\exp(2g_{0}\ell_{\rm g}) \geq \delta$.
	
	%where $\eta_{exci}$ is the excitation efficiency from the input power to the power at the upper laser level in the medium that is available in the form of inversion, $\eta_{extr}=\eta_{\rm B}\eta_{loss}$ represents the efficiency that the output laser power extracted from power at the upper laser level, $\etA_{\rm b}$ is the overlapping efficiency of the cavity mode and pump volume, and $\eta_{loss}$ depicts the overall loss during the energy transfer over the air and output coupling. $C$ represents the resonance oscillation threshold.
	
	%\begin{figure}[t]
	%    \centering
	%    \includegraphics[width=3.3in]{f-lossfactor}
	%    \caption{The loss factor in the procedure of resonant beam transmission.}
	%    \label{f:lossfactor}
	%\end{figure}
	
\begin{figure*}[t]
		\centering
		\includegraphics[width=4.5in]{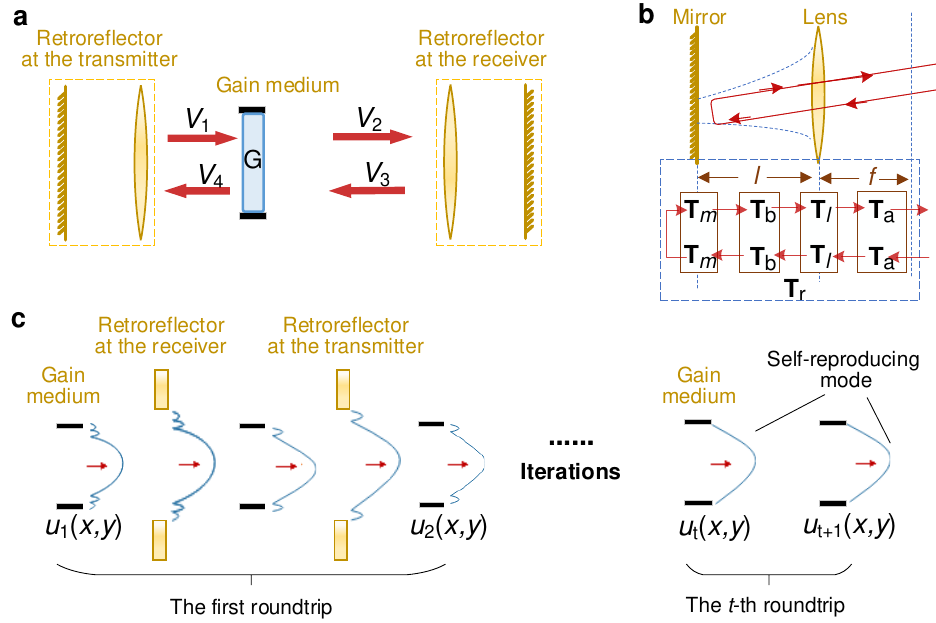}
		\caption{Loss factors, transition matrices of retroreflector, and Fox-Li algorithm. \textbf{a}, The loss factors in the procedure of resonant beam transmission. \textbf{b}, Transition matrices of field transfer through an FTCR. \textbf{c}, Fox-Li algorithm for iterative calculation of self-reproducing mode.}
		\label{f:lossfactor}
	\end{figure*}

	According to the resonator model used for the calculation of the output power~\cite{hodgson2005laser}, the traveling beam intensity is amplified through the gain medium, which is characterized by the small-signal gain coefficient $g_0$ varying with input power $P_{\rm in}$. On the other hand, the beam intensity is decreased due to diffraction losses (loss factors $V_1$-$V_4$), scattering, and absorption inside the medium, lenses, and mirrors (loss factor $V_{\rm S}$), and by output coupling (transmissivity of the retroreflector at the receiver is $1-R$). The output beam power is converted into electric power by a PV panel with the efficiency $\eta_{\rm pv}$. Thus, the electric power from the PV panel can be expressed as~\cite{hodgson2005laser}
	\begin{equation}
	\begin{aligned}
	P_{\rm out} = & \eta_{\rm pv} A_{\rm b} I_{\rm S} \\
	& \cdot \frac{(1-R) V_{2}}{1-R V_{2} V_{3}+\sqrt{R \eta_{\rm t}} \left[1 /\left(V_{1} V_{2} V_{\rm S}\right)-V_{\rm S}\right]} \\
	& \cdot \left[ g_{0} \ell_{\rm g}- \lvert \ln \sqrt{R V_{\rm S}^{2} \eta_{\rm t}}  \rvert \right],
	\end{aligned}
	\label{e:pout-methord}
	\end{equation}
	%{\color{black}what is $V_{\rm S}$?}
	where $\ell_{\rm g}$ and $I_{\rm S}$ are the length and the saturated light intensity of the gain medium, $A_{\rm b}$ is the cross section area of the resonant beam~\cite{beam1}. $V_1$,$V_2$,$V_3$, and $V_4$ indicate the transfer loss factor of the transmitter to the gain medium~\cite{hodgson2005laser}, the gain medium to the receiver, the receiver to the gain medium, and the gain medium to the transmitter, respectively, and $\eta_{\rm t} = V_1V_2V_3V_4$ represents the round trip loss factor, which we defined as over-the-air transmission efficiency, as shown in Fig.~\ref{f:lossfactor}a. %In our system, $\eta_{\rm t} = V$.
	
	The equation \eqref{e:pout-methord} can be reorganized as the equation \eqref{e:power} which indicates three power conversion steps including 1) from input power to the stored power in the gain medium, which is ready for stimulating the resonant beam with the efficiency $\eta_{\rm g}$; 2) the stored power in the gain medium to be extracted as the output beam power with the efficiency $\eta_{\rm e}$; and 3) the output beam power to electric power by photovoltaic conversion $\eta_{\rm pv}$. %, where the corresponding efficiencies are denoted by $\eta_{\rm g}$, $\eta_{\rm e}$, and $\eta_{\rm pv}$, respectively.
	Here, $\eta_{\rm pv}$ relies on the photovoltaic hardware, while $\eta_{\rm g}$ can be expressed as follows
	\begin{equation}
	\eta_{\rm g} =\frac{A_{\rm g}I_{\rm S}g_0\ell_{\rm g}}{P_{\rm in}},
	\end{equation}
	$\eta_{\rm e}$ is a function of $\eta_{\rm t}$ indicated by $F(\eta_{\rm t})$ in equation (1) as
	\begin{equation}
	\begin{aligned}
	\eta_{\rm e} &= F(\eta_{\rm t})\\& =  \frac{\eta_{\rm B}(1-R) V_2}{1-R V_2V_3+\sqrt{R \eta_{\rm t}} \left[1 /\left(V_{\rm S} V_1V_2\right)-V_{\rm S}\right]},
	\end{aligned}
	\end{equation}
	where $A_{\rm g}$ is the power stimulating (a.k.a, pumping) area of the gain medium, and $\eta_{\rm B} = A_{\rm b}/A_{\rm g}$ represents the overlapping efficiency. Thus, $\eta_{\rm g}$ relies on $g_0$ and $\ell_{\rm g}$, while $\eta_{\rm e}$ depends on $\eta_{\rm t}$. Moreover, in order to enable resonance, the input power to the gain medium should be above a threshold $P_{\rm th}$, which is depicted as
	\begin{equation}
	P_{\rm th} =\frac{\mid \ln \sqrt{R V_{\rm S}^{2} \eta_{\rm t}} \mid P_{\rm in}}{g_0\ell_{\rm g}}.
	\end{equation}

\subsection{Over-the-air transmission efficiency}
	To explore the over-the-air transmission efficiency $\eta_{\rm t}$, we analyze the electromagnetic field propagation inside the double-retroreflector cavity and obtain the self-reproducing mode.
	Based on the self-reproducing mode theory~\cite{hodgson2005laser}, after  being stimulated from the gain medium, the light wave will experience multiple bounces forth and back between the two retroreflectors. The light field mode is the complex value of field amplitude and phase, which can be self-reproducing after each round trip transfer until it becomes gradually stable. Once the mode reaches a steady state, the resonant beam for energy transfer is established. Owing to the double-retroreflector cavity, the stable self-reproducing mode can be established even when the receiver changes position.
	
	%We adopts here the diffraction theory of electromagnetic wave to
	%Thus, the path loss of the over-the-air energy transfer during the moving process can be obtained.
	Figure~\ref{f:transfer}c illustrates a three-dimensional $(x,y,z)$ coordinate system. We set $z = 0$ at the front surface of the gain medium, which is perpendicular to z-axis, i.e., on the $(x,y)$-plane. $u(x,y)$ is the field mode on the front surface of the gain medium. $\mathbf{u}$ is the vertical vector of all $u(x,y)$.
	The field transfer process is shown in Fig.~\ref{f:transfer}c, which can be formulated as a self-consistent equation as
	\begin{equation}
	\begin{aligned}
	\gamma \mathbf{u} &= \mathbf{T} \mathbf{u}, \\
	%\end{equation}
	%\begin{equation}
	\mathbf{u}=[u(x,y)]^T,
	\mathbf{T} &= \mathbf{T}_{\rm g}\mathbf{T}_{\rm r}\mathbf{T}_{\rm g}\mathbf{T}_{\rm s'}\mathbf{T}_{\rm r}\mathbf{T}_{\rm s}.
	\label{e:eigen}
	%\label{e:transfer}
	\end{aligned}
	\end{equation}
	%where
	$\gamma$ is the eigenvalue indicating the field change with one round-trip transmission, and the superscript $T$ denotes the transpose operation of a matrix. $\mathbf{T}_{\rm r}$, $\mathbf{T}_{\rm g}$, $\mathbf{T}_{\rm s}$, and $\mathbf{T}_{\rm s'}$ are the transition matrices of field transfer through inner-retroreflector, gain medium, free space from transmitter to receiver, and free space from receiver to transmitter, respectively. For example, $\mathbf{u}$ will be transferred to the field mode $\mathbf{T}_{\rm s}\mathbf{u}$ at the front surface of the retroreflector at the receiver by the transmitter-to-receiver field transfer depicted as $\mathbf{T}_{\rm s}$. Thus, $\mathbf{T}$ is the transition matrix of field transfer for one round trip, i.e.,  $\mathbf{T}_{\rm g}\mathbf{T}_{\rm r}\mathbf{T}_{\rm g}\mathbf{T}_{\rm s'}\mathbf{T}_{\rm r}\mathbf{T}_{\rm s}$ indicates the sequential transfer steps starting from the front surface of the gain medium and following the red-arrow-line loop.

$\mathbf{T}_{\rm g}$, $\mathbf{T}_{\rm s}$, and $\mathbf{T}_{\rm s'}$ are the transition matrices of field transfer through the gain medium, free space from transmitter to receiver, and free space from receiver to transmitter, respectively, which are expressed as
	\begin{equation}
	\begin{aligned}
	\mathbf{T}_{\rm g}\mathbf{u}=&[ u(x,y)\cdot G(x,y) ]^T  \\
	\mathbf{T}_{\rm s}\mathbf{u}=&[ \mathcal{F} ^{-1}\left\{\mathcal{F}\left\{u\left(x, y\right) \right\}\right. \\ &
	\cdot S(\nu_x,\nu_y,x_0+\Delta x,y_0+\Delta y)\\ & \cdot H(\nu_x,\nu_y,z_0+\Delta z) \}]^T \\
	\mathbf{T}_{\rm s'}\mathbf{u}=&[ \mathcal{F} ^{-1}\left\{\mathcal{F}\left\{u\left(x, y\right) \right\}\right. \\ &
	\cdot S(\nu_x,\nu_y,-(x_0+\Delta x),-(y_0+\Delta y))\\& \cdot H(\nu_x,\nu_y,z_0+\Delta z) \}]^T
	\end{aligned},
	\end{equation}
	where $\mathcal{F} $ and $\mathcal{F} ^{-1}$ denotes the two-dimensional fast Fourier transfer and inverse Fourier transfer; $(x_0,y_0,z_0)$ and $(\Delta x, \Delta y, \Delta z)$ in functions $H(\cdot)$ and $S(\cdot)$ represents the receiver's arbitrary initial position and the relative movement; thus, the new position  after movement is expressed as $(x_0 + \Delta x, y_0 + \Delta y, z_0 + \Delta z)$. %If the receiver moves along y-axis by $\Delta y$ and along z-axis by $\Delta z$ as in Fig.~\ref{f:transfer}c, $(0, \Delta y, \Delta z)$ is the receiver's coordinates after movement.
	
	%The gain medium will suffer gradients due to the temperature increment, which leading to the thermal lensing that the gain medium exhibits the function of beam focusing similar as a convex lens. Thus, here the gain medium is equivalent to a lens with focal length $f_{t}$, which is depicted as
	%\begin{equation}
	%    G(x, y)= \left\{\begin{aligned}
	%\exp{\left[-i\frac{\pi}{\lambda f_t}(x^2+y^2)\right]}&, x^{2}+y^{2} \leq r_{\rm g}^{2} %\\ 0&, \text { else }
	%\end{aligned}\right.,
	%\end{equation}
	%where $r_{\rm g}$ is the radius of the gain medium.
	
	Here the gain medium is equivalent to  an aperture $G(x,y)$ which is expressed as an indication function to depict the area within the aperture by~1 and the area outside the aperture by~0; that is
	\begin{equation}
	G(x, y)= \left\{\begin{aligned}
	1&, x^{2}+y^{2} \leq r_{\rm g}^{2} \\ 0&, \text { else }
	\end{aligned}\right.,
	\end{equation}
	where $r_{\rm g}$ is the radius of the gain medium.
	
	$H(\nu_x,\nu_y,d_{\rm L})$ is the free space propagation kernel of the field mode as~\cite{FFT}
	\begin{equation}
	H(\nu_x,\nu_y,d_{\rm L})=\exp{\left[ i\frac{2\pi}{\lambda}d_{\rm L}\sqrt{1-(\lambda \nu_x)^2-(\lambda \nu_y)^2}\right]},
	\end{equation}
	where $\nu_x$ and $\nu_y$ represent the spatial frequency coordinates, $d_{\rm L}$ is the propagation distance, $\lambda$ is the wavelength, and $i = \sqrt{-1}$. $\mathbf{T}_{\rm a}\mathbf{u}$ is actually the Rayleigh-Sommerfeld scalar diffraction integral with fast Fourier transform method which depicts the optical fields propagation and $H(\cdot)$ is the system transfer function of free-space propagation~\cite{FFT,Koechner2013Solid,hodgson2005laser}.
	
	After the receiver's movement, the retroreflector at the receiver is offset with respect to the optical axis of the previous plane in our system. $S(\nu_x,\nu_y,s_x,s_y)$ indicates the free-space propagation shift in the spatial frequency domain given by~\cite{shift}
	\begin{equation}
	S(\nu_x,\nu_y,s_x,s_y) = \exp{\left[i2\pi(\nu_x{s}_x+\nu_y{s}_y)\right]},
	\end{equation}
	where $s_x$ and $s_y$ are relative shift distances along x-axis and y-axis of the coordinates system between the latter plane and former plane with the field propagation.
	
	%\begin{figure}[t]
	%    \centering
	%    \includegraphics[width=1.3in]{reflectorMatrx}
	%    \caption{Transition matrix of field transfer through a retroreflector.}
	%    \label{f:reflector}
	%\end{figure}

	$\mathbf{T}_{\rm r} = \mathbf{T}_{\rm a}\mathbf{T}_{l}\mathbf{T}_{\rm b}\mathbf{T}_{\rm m}\mathbf{T}_{\rm m}\mathbf{T}_{\rm b}\mathbf{T}_{l}\mathbf{T}_{\rm a}$ denotes the transition matrix of the retroreflector as in Fig.~\ref{f:lossfactor}b. We choose the focal plane at the mirror-opposite side of the lens as the original plane to depict the field transfer calculation of the retroreflector. $\mathbf{T}_{\rm m}$ is the transition matrix for field transfer through a mirror (equivalent to an aperture), which is generally expressed
	%as an unity matrix, i.e.,
	as an indication function to depict the reflective area by 1 and non-reflective area by 0. $\mathbf{T}_{l}$, $\mathbf{T}_{\rm b}$, and $\mathbf{T}_{\rm a}$ are the transition matrices for field transfer through a convex lens, the interval between the lens and mirror, and interval between the lens and the focal plane. The corresponding field transfer operations are defined as
	\begin{equation}
	\begin{aligned}
	&\mathbf{T}_{\rm m}\mathbf{u}=[ u(x,y)\cdot M(x,y) ]^T \\
	&\mathbf{T}_{\rm l}\mathbf{u}=[ u(x,y) \cdot L(x,y)]^T \\
	&\mathbf{T}_{\rm b}\mathbf{u}=[ \mathcal{F} ^{-1}\left\{\mathcal{F}\left\{u\left(x, y\right) \right\}  \cdot H(\nu_x,\nu_y,l) \right\}]^T \\
	&\mathbf{T}_{\rm a}\mathbf{u}=[ \mathcal{F} ^{-1}\left\{\mathcal{F}\left\{u\left(x, y\right) \right\}\right.  \cdot H(\nu_x,\nu_y,f) \}]^T
	\end{aligned},
	\end{equation}
	where $M(x,y)$ is the indication function and $L(x,y)$ represents the phase changes after the field passes through a convex lens defined as
	\begin{equation}
	M(x, y)= \left\{\begin{aligned}
	1&, x^{2}+y^{2} \leq r^{2}, \\ 0&, \text { else },
	\end{aligned}\right.
	\end{equation}
	and
	\begin{equation}
	L(x, y)= \left\{\begin{aligned}
	&\exp{\left[-i\frac{\pi}{\lambda f}(x^2+y^2)\right]}, x^{2}+y^{2} \leq r^{2}, \\ & 0, \text { else },
	\end{aligned}\right.
	\end{equation}
	where $r$ is the radius of the retroreflector front surface, $l$ is the interval between the flat mirror and lens in the retroreflector, $f$ is the focal length of the lens. %, and $\lambda$ indicates the wave length of the resonant beam.

	%\begin{figure}[t]
	%    \centering
	%    \includegraphics[width=3.5in]{f-Fox-Li}
	%    \caption{The Fox-Li algorithm in the proposed system.}
	%    \label{f:foxli}
	%\end{figure}
	After many reflections inside the double-retroreflector cavity, the resonant beam field distribution will not vary after transition and reach a steady-state field distribution, i.e., the self-reproducing mode. To solve equation (2) and obtain the self-reproducing mode, we adopted the Fox-Li algorithm to simulate the beam oscillation process inside the resonator with iterative calculation as in Fig.~\ref{f:lossfactor}c~\cite{FoxLi}.
	
	The process that the resonant beam reflected back and forth in the SSLR is equivalent to the beam propagation through a serial optical system (i.e., the lenses, the mirrors, the gain medium, and the space propagation between these deceives). We assume an arbitrary initial field at the gain medium and then compute the field hereafter one round trip transmission as a result of the first transition. The newly calculated field after normalization is then used to compute the field at the gain medium after the second transition.
	Such computation will be repeated for successive transitions until a steady-state is reached.
	Thus,
	%set the original field on the gain medium as $U(x,y) = 1$,
	we can obtain the self-reproducing mode through iteratively computing equation (2) using this Fox-Li algorithm. The iteration can be stopped % at $t$ iteration times
	as the field becomes  stable and eventually the self-reproducing mode is obtained. If the receiver is shifted by arbitrary $(\Delta x, \Delta y, \Delta z)$ relative to its original position $(x_0,y_0,z_0)$, only $\mathbf{T}_{\rm s}$ and $\mathbf{T}_{\rm s'}$ need to be updated, while $\mathbf{T}_{\rm g}$ and $\mathbf{T}_{\rm r}$ are invariant to movement.
	%where $U(x,y,z)$ is the {\color{black}field profile, i.e., the complex value of field amplitude and phase at position $(x,y,z)$; $d$ is the distance from the origin along z-axis; $\mathbf{u}$ is the vertical vector of $U(x,y,z=d)$ for all positions on the $(x,y)$-plane at $z=d$. We set $z = 0$ at the front surface of the gain medium, which is perpendicular to z-axis. Thus, $U(x,y,0)$ depicts the field profile on the front surface of the gain medium. $\mathbf{T}$ is the transition matrix of field transfer for one round trip and $\mathbf{T}_{\rm g}\mathbf{T}_c\mathbf{T}_{\rm g}\mathbf{T}_{\rm s'}\mathbf{T}_c\mathbf{T}_{\rm s}$ indicates the sequential transfer steps. }
	
	The eigensolution of the above self-consistent equation \eqref{e:eigen} can be obtained by iterative calculation~\cite{hodgson2005laser}. The field mode will gradually reach a steady state, i.e., achieving a self-reproducing mode, as the iterative index $t$ increases to a certain value. %Thus, we can obtain
	%the over-the-air transmission efficiency and
	%the field profile distribution inside the cavity for different receiver's positions.
	The eigensolution %$\mathbf{u}_e$
	of the above equation describes the field amplitude distribution of the steady-state resonant beam, which is self-reproducing after one round trip transfer. The corresponding eigenvalue $\gamma$ depicts the field mode change after one round trip transfer due to diffraction loss through the over-the-air transmission, and $|\gamma|^2$ is the field intensity change. Thus, the over-the-air transmission efficiency can be calculated as~\cite{siegmanlaser}
	\begin{equation}
	\eta_{\rm t} = |\gamma|^2 =
	%\lim_{t\to \infty}\frac{||\bold{u_{t+1}}||}{||\bold{u_{t}}||}
	\lim_{t\to \infty}\frac{\iint_{(x,y)}|u_{t+1}(x,y)|^2}{\iint_{(x,y)}|u_{t}(x,y)|^2},
	%\lim_{t\to \infty}\frac{\iint_{(x,y)}|U_{t+1}(x,y,z)|^2}{\iint_{(x,y)}|U_{t}(x,y,z)|^2},
	\end{equation}
	where $u_{t}(x,y)=u(x,y)$ for the $t$-th iteration. Even though the front surface of the gain medium is selected as an example in the above analysis, any other surface along the round trip transfer in Fig.~\ref{f:transfer}c can also be chosen, which leads to the same $\eta_{\rm t}$ using the same analysis.
	
	%Thus, we can obtain
	%the over-the-air transmission efficiency and
	%the field profile distribution inside the cavity for different receiver's positions.
	
	Based on the above eigenmode analysis, the resonant beam in the double-retroreflector cavity with arbitrary movement can be depicted in terms of the self-reproducing mode and the over-the-air transmission efficiency. The double-retroreflector design guarantees the steady state of the self-reproducing mode in a certain movement region without tracking control. Readers should be noticed that although we investigate the over-the-air transmission efficiency here, many other factors such as the pump efficiency, the absorbing efficiency of the gain medium, the output laser transmission efficiency, and the PV conversion efficiency also affect the overall WPT efficiency; and we will discuss them in Section~\ref{sec:discu}.

	%. Hence this design automatically guarantees the stability of the resonant beam %for mobile WPT.
	
	%Thus, we can obtain the solutions for self-reproducing mode of the cavity in the presentd system for the movement with different shift and angle.
	
	%\begin{itemize}
	%\item The field transfer within the doubled cat's eye cavity can be expressed as the following self-consistent equation.
	%\item The round-trip transfer loss $\gamma$ can be obtained with the above equation, varying with the mobility factor: distance and angle.
	%\end{itemize}
	
	%\subsection{Power output}
	%If the retroreflector at the receiver is partially reflected, as $R_r<100\%$, the  resonant beam will partly pass through the retroreflector to form the output beam, i.e, laser beam. The output beam incidents on the photovoltaics panel to generate electric power.
	
	%\begin{itemize}
	%\item The field transfer within the doubled cat's eye cavity can be expressed as the following self-consistent equation.
	%\item Power conversion with the PV panel.
	%\end{itemize}
	
	%\section{Experiments and Simulations{\color{black}(12/23)}}

	\section{Calculation and Experiment}
	\begin{figure}[t]
		\centering
		\includegraphics[width=3.4in]{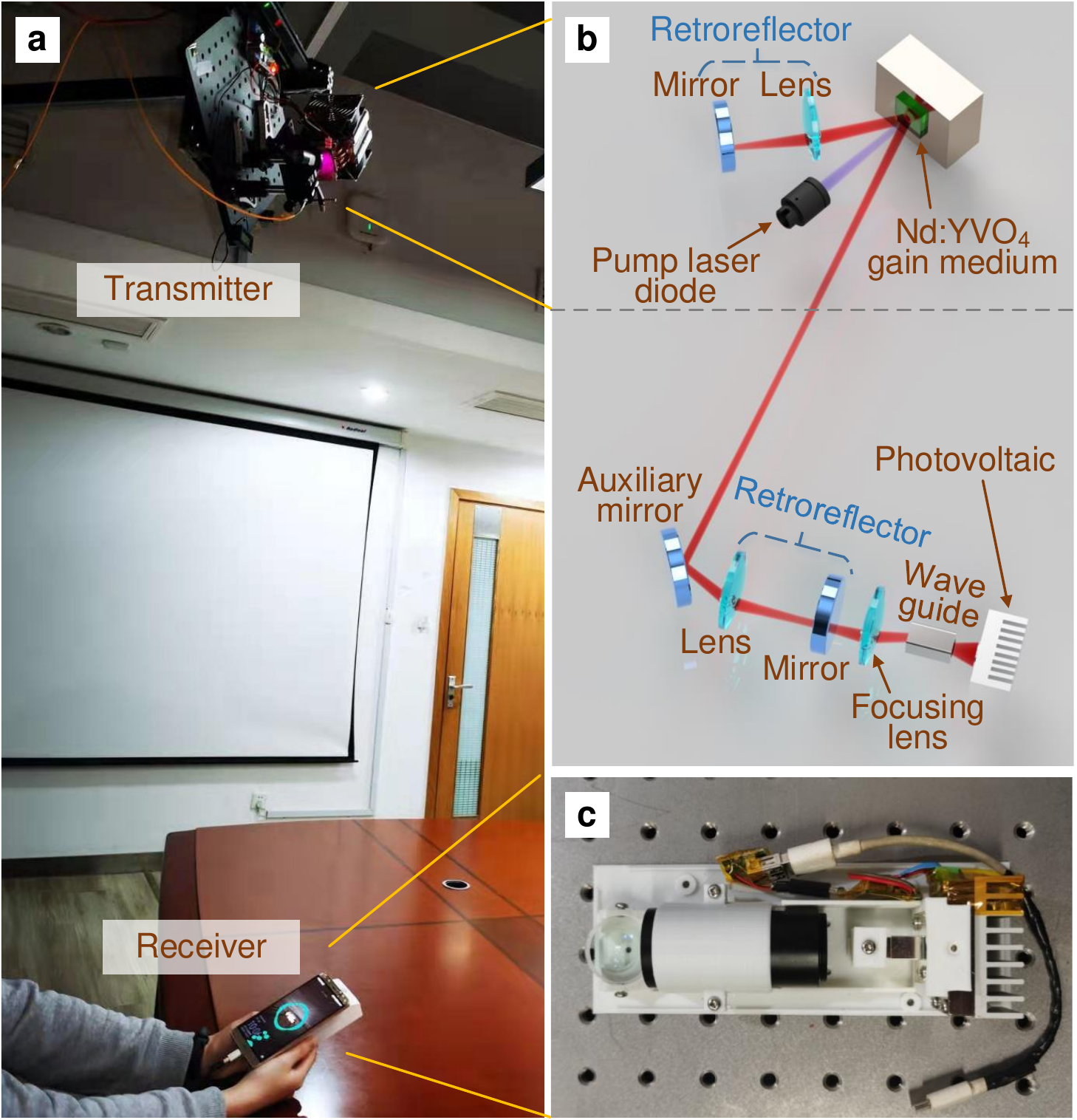}
		\caption{Experimental verification.% of resonant beam charging with the double-retrorelector cavity.
			 ~\textbf{a}, Experiment of charging a smartphone. \textbf{b}, Experimental components. \textbf{c}, The receiver with the size $16.7 \times 6.0 \times 3.6$~cm$^3$.}
		\label{f:experiment}
	\end{figure}

	Here, we demonstrate the theoretical analysis and the experiment of the resonant beam charging system.
	Previously, optical systems have been used to illustrate the physics of resonant beam for non-mobility WPT~\cite{WWang2018}. Our work here differs in that we demonstrate a mobile WPT scheme.
	
\subsection{Measurement setup and parameters}

	Figure~\ref{f:experiment} shows the experiment setup. The transmitter consists of an FTCR and an Nd:YVO$_{\text{4}}$ gain medium. A pump laser diode with $878$-nm output  feeds power to the gain medium to stimulate $1064$-nm
	infrared resonant beam. %{\color{black} The feeding power to the gain medium is $40$ W.}
	The receiver consists of an FTCR with $90\boldsymbol{\%}$ reflectivity, a focusing lens, a waveguide, and a PV panel. The focusing lens directed the beam at an arbitrary direction into the waveguide. The waveguide homogenizes the beam intensity to increase the consequent PV conversion efficiency.
	Finally, The output light from the waveguide is converted by the PV panel into electricity. The radii of lenses and mirrors in the retroreflectors at both the transmitter and the receiver are $7$~mm.
	The PV panel size is $10\times6$~mm$^2$. %There is also a heat sink is mounted onto the photovoltaic cell for heat dissipation. %The above components are with compact size and are capable of being embedded in mobile products. In addition to the experiment, we fabricate a portable receiver to demonstrate mobile charging applications.
	A direct current to direct current~(DC-DC) converter is used to convert the high-voltage and variable PV output to a stable $5\mbox{-V}$ charging voltage which is acceptable for most electronic devices~\cite{WWang2018,zhang2019}.

\begin{table}[t]
		\centering
		\caption{Parameters in theoretical calculation}
		\vspace{.7em}
		\begin{tabular}{ccc}
			\hline
			\textbf{Parameter}&\textbf{Symbol}&\textbf{Value}\\
			\hline
			%\text{Resonant beam wavelength}&$\lambda$&$1$ $\mu$m\\
			\text{Gain medium radius}&$r_{\rm g}$&$2.8$ mm\\
			\text{Radius of lens and mirror}&$r$&$7.0$ mm\\
			\text{Retroreflector focal length}&$f$ & $50.4$ mm\\
			\text{Retroreflector inner interval}&$l$ & $52.0$ mm\\
			\text{Input power to gain medium}&$P_{\rm in}$&$37.3$ W\\
			\text{Medium saturated intensity}&$I_s$&$1.26\times10^{7}$ W/m$^2$\\
			\text{Output retroreflector reflectivity}&$R$&$90\%$\\
			\text{Loss factor by optical components}&$V_s$&$0.88$\\
			\text{Excitation efficiency}&$\eta_{\rm g}$&$0.72$\\
			\text{Photovoltaic efficiency}&$\eta_{\rm pv}$&$0.12$\\
			\hline
			\label{t:systemPara}
		\end{tabular}
	\end{table}

\begin{figure}
	\centering
	\includegraphics[width=3.5in]{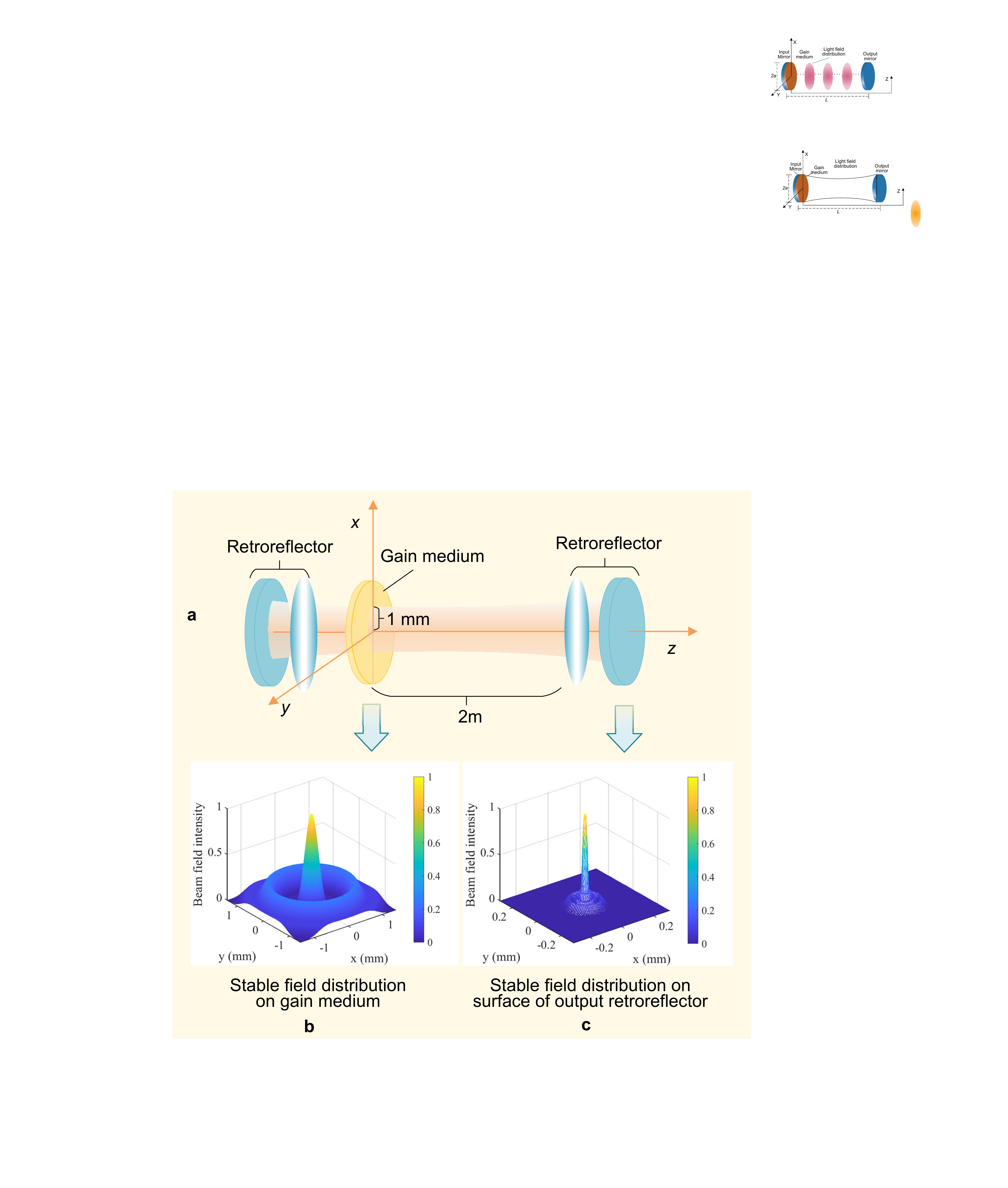}
	\caption{Stable field distribution in the FTCR-based resonator.~\textbf{a}, Simulation model of an SSLR. \textbf{b}, Stable field distribution on gain medium. \textbf{c}, Stable field distribution on surface of output retroreflector.}
	\label{f:results}
\end{figure}

A $12 \times 12$~mm Nd:YVO$_4$ gain medium is adopted at the transmitter. A steel disc with a hole ($2.8$-mm radius) is attached to the surface of the gain medium to fix the medium onto the cooling module. All mirrors and lenses in the retroreflectors are attached to an  aperture with a radius of $7$~mm. At the receiver, the waveguide is coated with silver so that most light energy can be guided to the PV panel. The FTCRs at the transmitter and receiver share the same configuration. The focal length of the lens in the retroreflector is $50$~mm ($\pm 1\%$). The interval between the lens principal plane and the mirror is $52$~mm (the principal plane of the lens is closer to the mirror than the center of the lens,  and the interval between the lens principle plane and the the lens center is calculated to be $0.9$~mm). The input power to the gain medium is $37.3$~W. The reflectivity of the mirror in the FTCR at the transmitter is near $1$. However, the reflectivity of the mirror at the receiver is $90\%$, which means $10\%$ beam power can pass through the mirror. All the experiment results are obtained as the average of more than 5 measurements to eliminate any possible random factors. We calculated the field distribution, the output beam power, the output electrical power, and the over-the-air transmission efficiency using the parameters estimated from the experimental setup.  We started each calculation with the initial plane wave field values and waited until a steady state was reached before recording the relevant results. Table~\ref{t:systemPara} contains the parameters in theoretical calculation.

\subsection{Theoretical and experimental results}
	In Fig.~\ref{f:results}, we obtain the eigensolution of the above self-consistent equation \eqref{e:eigen} by numerical calculation and show the stable field distribution inside the SSLR. The beam field intensity is spatially-concentrated with less than 1-mm radius along the 2-m axis between the two FTCRs, which forms a narrow beam shape as observed in the experiments.  %intensity distribution in %the output beam power and the over-the-air transmission efficiency in Fig.~\ref{f:moving1} and in Fig.~\ref{f:moving2} .
		
\begin{figure}
	\centering
	\includegraphics[width=3in]{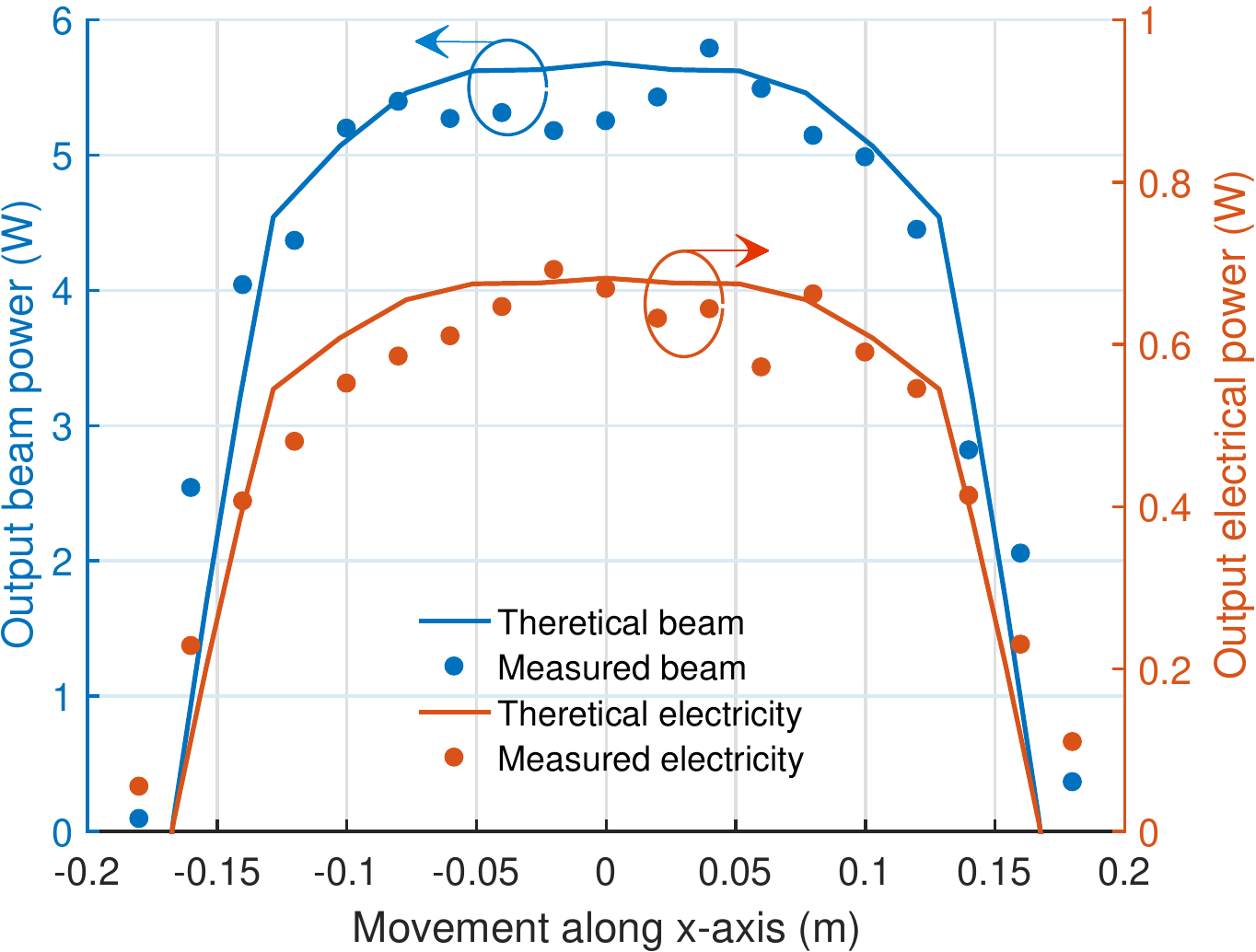}
	\caption{Output beam power and electrical power vs. receiver's movement along x-axis.}
	\label{f:xpower}
\end{figure}

\begin{figure}
	\centering
	\includegraphics[width=3in]{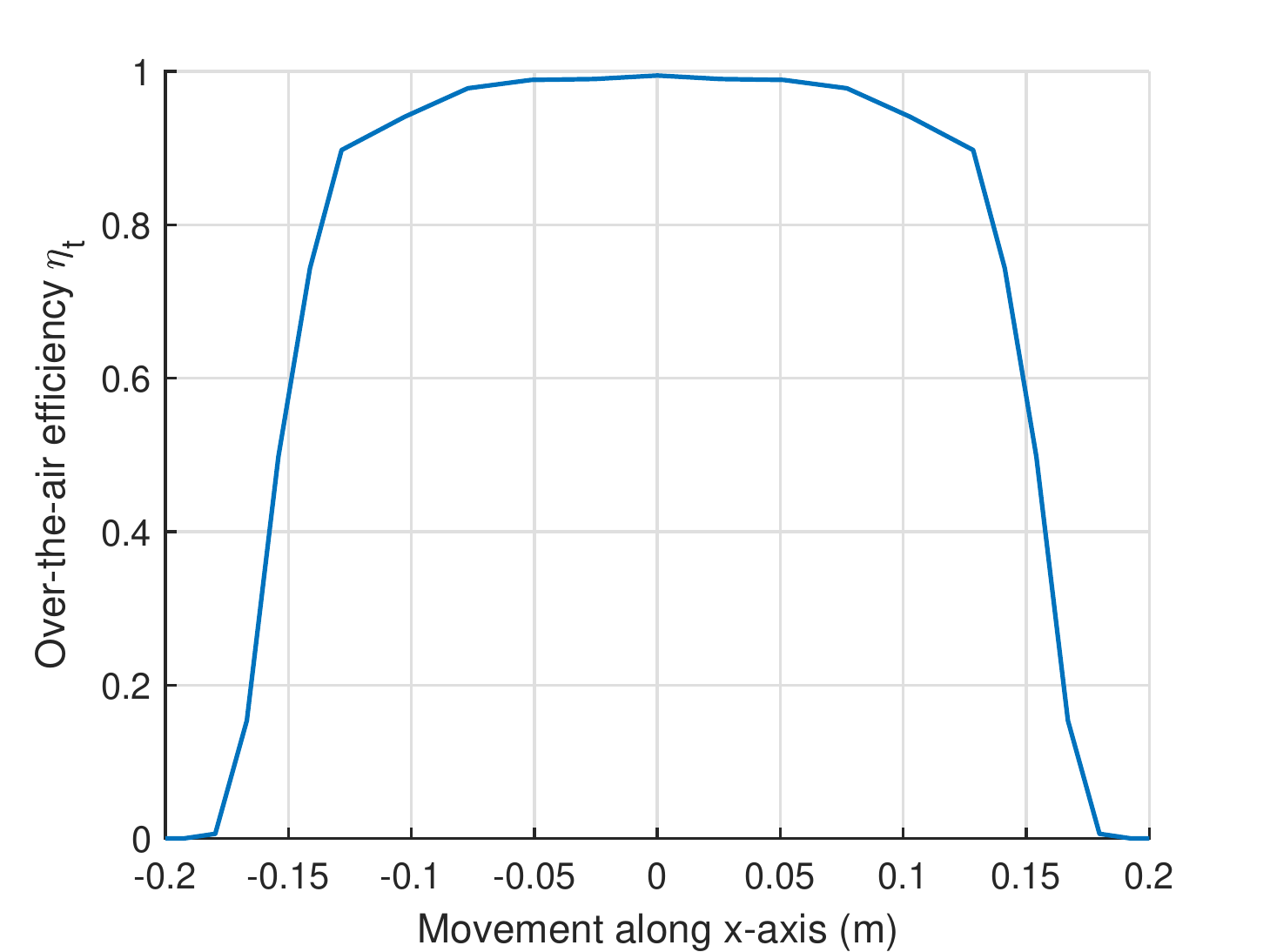}
	\caption{Over-the-air transmission efficiency $\eta_{\rm t}$ vs. receiver movement along x-axis $\Delta x$ from $(x_0=0,y_0=0,z_0=2~\mbox{m})$.}
	\label{f:xeff}
\end{figure}

	In Fig.~\ref{f:xpower}, for 2-m vertical transmission distance, the output beam power and the output electrical power are measured and calculated. The transmitter is located at the origin $(0,0,0)$ and the receiver is shifted on the $(x,y)$-plane along x-axis between $(-\Delta x,0,2~\text{m})$ and $(\Delta x,0,2~\text{m})$. When the shifting distance $\Delta x < 10$ cm, the output beam power is above $5$~W, the output electrical power is around $0.6$~W, and the over-the-air transmission efficiency remains nearly $100\boldsymbol{\%}$ as in Fig.~\ref{f:xeff}. When $\Delta x > 10$~cm, the output beam power, the output electrical power, and the transmission efficiency drop down to zeros gradually. Thus, $-10$ cm $< \Delta x < 10$~cm at $z=2$~m from the transmitter is equivalent to around $6^{\circ}$ field of view. Besides, we can observe that the maximum horizontal moving range (along the $x$-axis) where the power can be received is up to $\pm 18$~cm.

\begin{figure}
	\centering
	\includegraphics[width=3in]{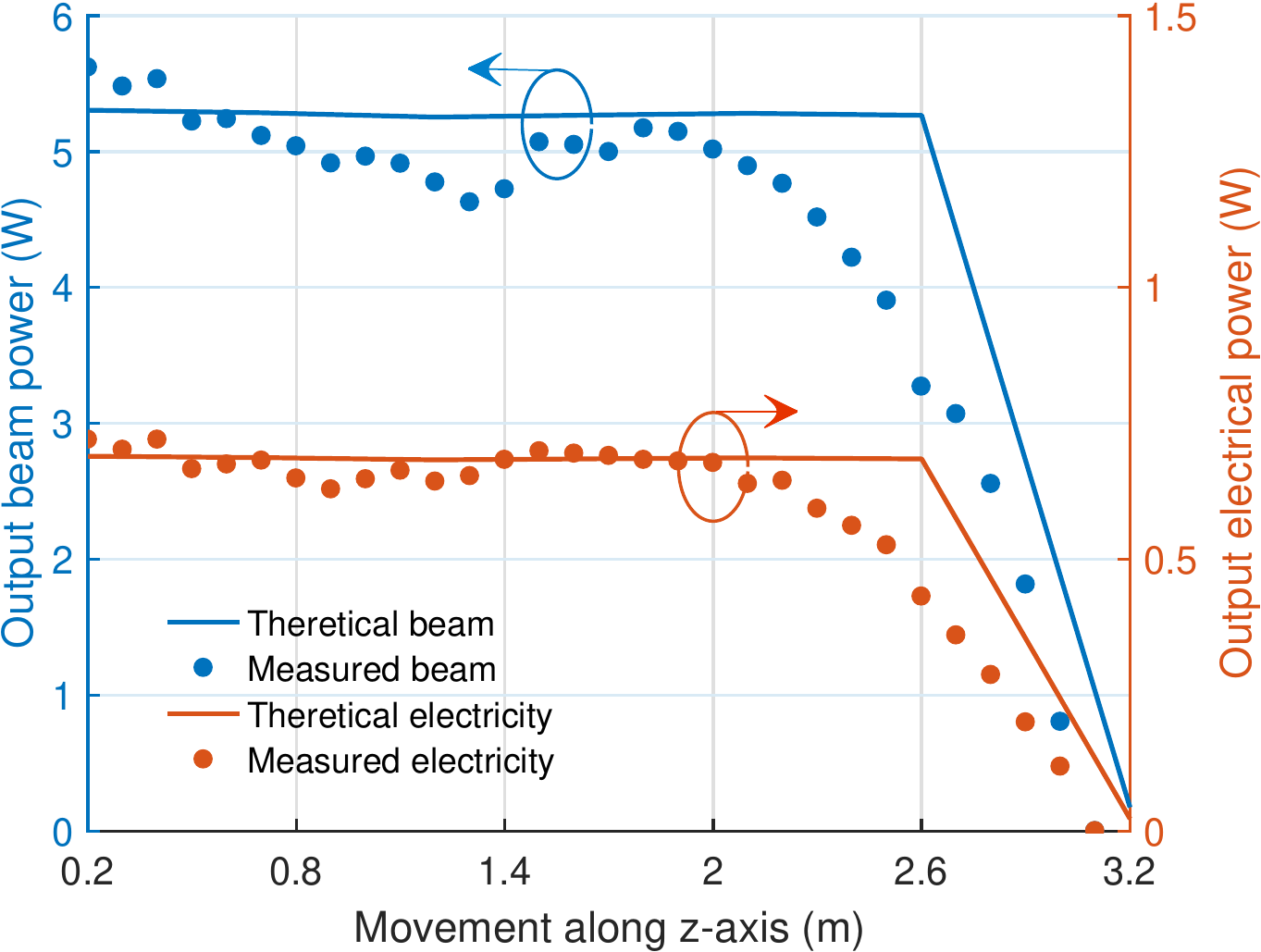}
	\caption{Output beam power and electrical power vs. receiver's movement along z-axis.}
	\label{f:ypower}
\end{figure}

\begin{figure}
	\centering
	\includegraphics[width=3in]{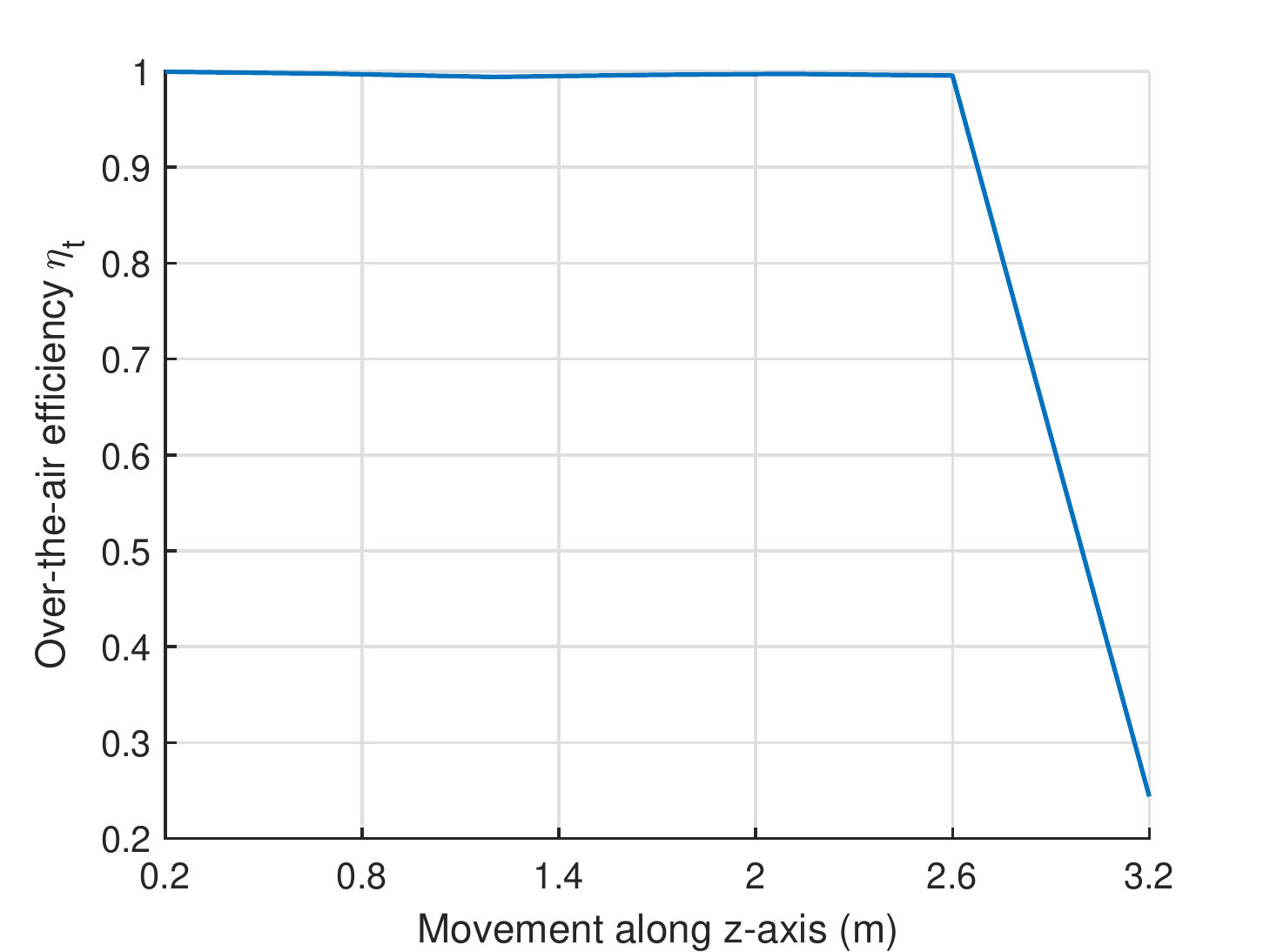}
	\caption{$\eta_{\rm t}$ vs. receiver movement along z-axis $\Delta z$ from  $(x_0=0,y_0=0,z_0=0)$.}
	\label{f:yeff}
\end{figure}

	In Fig.~\ref{f:ypower}, %the output beam power and the output electrical power are shown.
	the transmitter is located at $(0,0,0)$ and the receiver is moved along $z$-axis as $(0,0, \Delta z)$. When $0.2$~m~$<\Delta z < 2$~m, the output beam power is around 5~W, the output electrical power is above $0.6$~W, and the over-the-air transmission efficiency remains nearly $100\boldsymbol{\%}$ as in Fig.~\ref{f:yeff}. When $\Delta z > 2$~m, the output optical power, the output electrical power, and the transmission efficiency start turning down to zeros gradually. We can observe that the maximum vertical moving distance (along the $z$-axis) is up to $3$~m. The maximum transmission distance depends on the stability of the SSLR. \cite{xiongretro} gives the boundary of the transmission distance for such a symmetric SSLR, i.e., $d<4f_{\rm RR}$, while the real transmission distance also depends on the losses and the gain.

	The results of the numerical calculation are generally in good agreement with the corresponding measurement. Therefore, such a theoretical model can indeed be used to demonstrate the power transfer in the resonant beam charging system as predicted from the self-reproducing mode theory.

	\section{Discussion}
	\label{sec:discu}
	{\textit{Mobility}}:
	We demonstrate the above experimental system for charging a smartphone in a mobile operation. From the above theoretical and experimental results, the WPT system can continuously supply above 0.5-W electrical power %{\color{black}(measured in the above experiments as shown in Entended Data Fig. XX)}
	%(e.g., the photovoltaic efficiency $10\boldsymbol\%$ and the output beam power $5$ W)}
	%with {\color{black}near $100\boldsymbol\%$} transmission efficiency
	for a compact-size moving receiver in the coverage of 2-m distance and $6^{\circ}$ field of view from the transmitter. %, i.e., in a 2-square-meter area.
	A typical power supply for a smartphone is 5~V and 1~A (i.e., 5~W). Assume the smartphone needs to be fully charged once per day, and each charging duration is $1$ hour. Thus, its energy consumption is 5~Wh per day. While using the demonstrated resonant beam charging system, the smartphone can be completely charged with $0.5\mbox{-W}$ charging power in 10 hours. Although the charging time is expanded, users no longer need to carry a power cable.
	Since the charging even occurs in mobile operation, the prospective charging time could be more than 10~hours per day, e.g., in the scenario of charging the smartphone in the area of an office table. Thus, this WPT system enables unlimited battery life for this smartphone, where charging via a power cable is no longer needed. It facilitates imaginative applications in the future such as a mobile charging network that delivers energy to anything, anywhere, and anytime similar to mobile Internet.

	%{\color{black}(Efficiency)
	%{\color{black}\textbf{\textit{Efficiency:}}}
	{\textit{Efficiency}}:
	Improvement of the power transfer efficiency from the transmitter to the receiver can be achieved by enhancing the efficiency of the power feeding to the gain medium at the transmitter, the over-the-air beam transmission, and the beam-to-electricity power conversion at the receiver. At the transmitter, the gain medium in our experiment is solid-state material. To improve the conversion efficiency of the gain medium, high efficient semiconductor gain medium could be used~\cite{Vecsel,vecsel2}. To increase the over-the-air transmission efficiency, adaptive mode control schemes, e.g., an adjustable optical modulator can be adopted~\cite{opticalm}. At the receiver, we chose an off-the-shelf PV panel in our experiment. If the PV panel is optimized, e.g.,
	designing for the specific wavelength, the output beam conversion efficiency can be enhanced~\cite{PV1,PV2,PV3,PV4,PV5}. Besides, although in this work the output mirror's reflectivity $R$ is $90\%$, changing this parameter will also affect the overall transmission efficiency. Thus, to obtain a maximum output, one should find an optimum $R$.

	%{\color{blackv}(Safety)}
	%{\color{black}\textbf{\textit{Safety:}}}
\begin{figure}[t]
    \centering
    \includegraphics[width=2.3in]{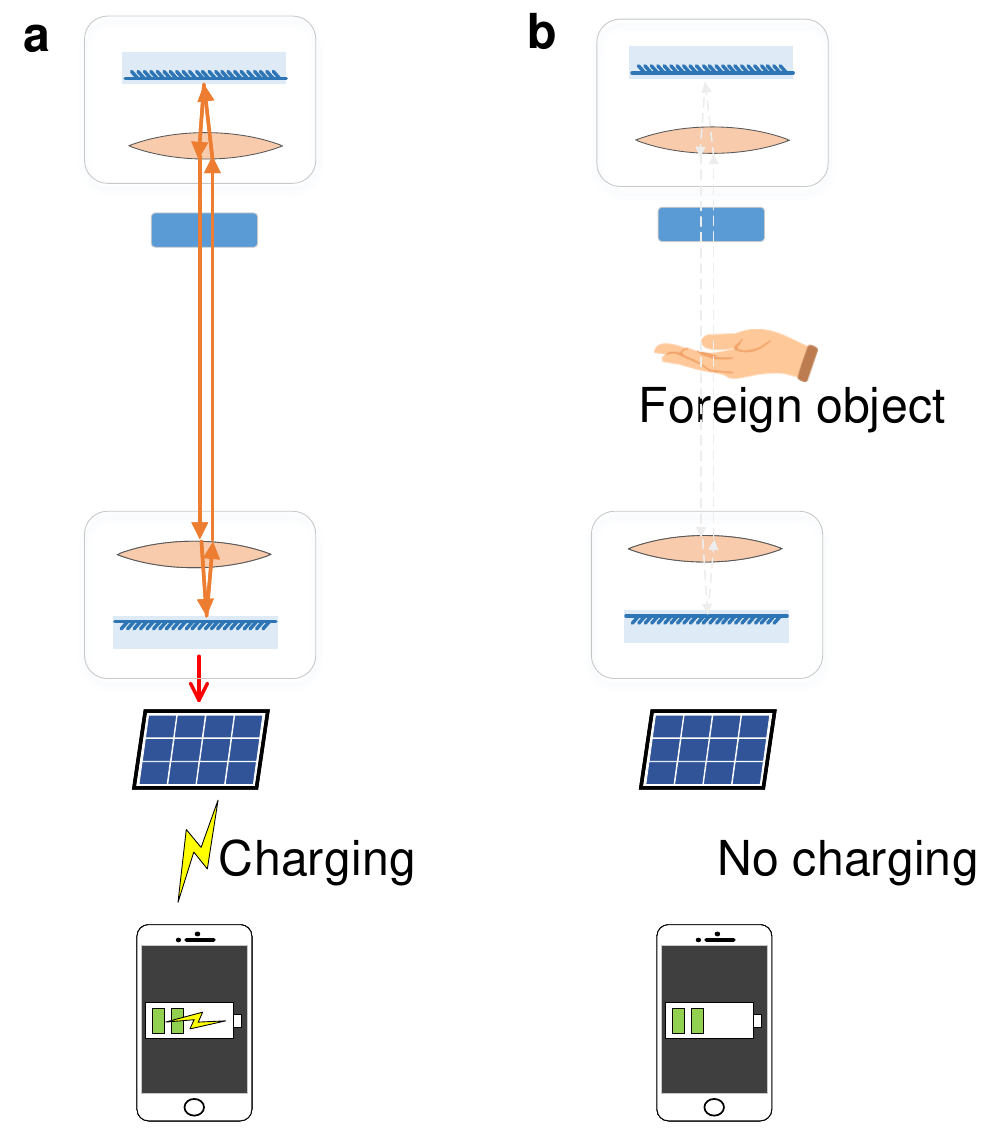}
    \caption{Comparison of the unobstructed and obstructed scenarios. \textbf{a}, The resonant beam is unobstructed. \textbf{b}, The resonant beam is obstructed.}
    \label{f:safety}
\end{figure}

\begin{table}[t]
	\centering
	\caption{Safety evaluation for foreign object invasion.}
	\vspace{.7em}
	\begin{tabular}{ccc}
		\hline
		\textbf{Parameter}&\textbf{Symbol}&\textbf{Value}\\
		\hline
		\text{Cavity length}&$L$&$2$ m \\
		\text{Light speed}&$c$ & $3\times10^8$ m/s \\
		\text{Storage time}&$t_s$&$6.67\times10^{-9}$ s\\
		%\text{Transmission efficiency}&$\eta_{\rm t}$&$90\%$ \\
		\text{Output beam power}&$P_{\rm o}$&$5$ W \\
		\text{Intra-cavity stored power}&$P_{\rm s}$&$50$ W\\
		\text{Beam cross-sectional area}&$A_r$&$1.54\times10^{-4}$ m$^2$\\
		\textbf{Beam radiant exposure}&$E_r$&$0.0022$ J/m$^2$\\
		\hline
		\textbf{Maximum permissible exposure}&$E_m$&$1000$ J/m$^2$\\
		\hline
		\label{t:Safetycal}
	\end{tabular}
\end{table}

	{\textit{Safety}}:
	As in Fig.~\ref{f:safety}, if a foreign object, such as a hand, is placed between the transmitter and the receiver, i.e., obstructs the line of sight (LOS) between the two retroreflectors, the round-trip transfer loss will suddenly increase beyond the gain, which breaks the oscillation condition; thus, the resonance will be cut off. Upon the foreign object's obstruction, the radiant exposure of resonant beam in the above experimental system is calculated as $0.0022$~J$/$m$^2$, which is less than the maximum permissible exposure $1000$ J$/$m$^2$ in the IEC$60825$-1:2014 laser safety standard~\cite{Safety2014}. The safety of foreign objects which gradually invade into the resonant beam has been quantitatively analyzed by numerical simulation in~\cite{a211015.04}. As a foreign object enters the resonator, the modes of the resonant beam change relatively with the progress of the invasion. As the modes with larger radius disappear, the power transits to the inner modes, keeping the electromagnetic field
	leakage in an allowable range.  In our mobile resonant beam charging  experiments, we also verified the invasion safety. The output laser is still safe for users, as it is directed into a waveguide and absorbed by the PV panel.
	
	When the resonant beam is cut off by a foreign object, the maximum beam radiant exposure $E_{\rm r}$, i.e., the radiant energy per unit area, to the foreign object can be calculated based on the circulating power model~\cite{Koechner2013Solid}. $P_{\rm s}$ is the intra-cavity resonant beam power, which can be depicted as:
	\begin{equation}
	P_{\rm s} = \frac{P_{\rm o}}{1 - R},
	\label{e:intrapower}
	\end{equation}
	where $P_{\rm o}$ is the output beam power, and $R$ is the reflectivity of the retroreflector at the receiver. The power storage time $t_{\rm s}$ in the cavity can be calculated based on the cavity length (i.e. transmission distance) $L$ and the light speed $c$; that is
	\begin{equation}
	t_{\rm s} = \frac{L}{c},
	\label{e:storetime}
	\end{equation}
	Thus, the beam radiant exposure $E_r$ can be depicted as
	\begin{equation}
	E_{\rm r} = \frac{P_{\rm s} t_{\rm s}}{A_{\rm r}},
	\label{e:exposure}
	\end{equation}
	where $A_{\rm r}$ is the cross-sectional area of the resonant beam.
	We have performed the calculation for the experimental system and obtained $E_{\rm r}=0.0022 {\text{ J}/\text{m}^2}$ as in Table \ref{t:Safetycal}, which is below the MPE to skin of laser radiation safety regulation $E_{\rm m}=1000 {\text{ J}/\text{m}^2}$ in the IEC $60825$-1:$2014$ standard.
	In order to improve safety, the radius and reflectivity of the retroreflector can be adjusted to reduce radiant exposure. %The safety of the presentd scheme has been verified theoretically and experimentally.
	
	%{\color{black}(Extendibility)}
	%{\color{black}\textbf{\textit{Extendibility:}}}
	{\textit{Extensibility}}:
	Our work adopts an infrared wavelength, i.e. $1064$ nm, as an example to verify the theory with the experiments. The principle of the resonant beam in the SSLR could also be applied to other wavelengths, e.g., radio frequency~\cite{NS1,NS2}, which may be a new direction in the future study of mobile WPT.
	
	\section{Conclusion}
	In conclusion, we presented both a theoretical study and an experiment on a self-aligned resonant beam in the double-retroreflector-based spatially separated laser resonator~(SSLR) for mobile and over-the-air wireless charging. The spatially-concentrated field distribution of the resonant beam resulted in high-efficiency transmission over a long distance. Particularly, the SSLR based on focal telecentric cat's eye retroreflectors enabled automatically charging a compact-size moving receiver without positioning/tracking control. Experiment results were in good agreement with the theoretical calculation. This work provides guidelines for theoretical and experimental research in the future. Several improvement points such as expanding the moving range, increasing the total efficiency, and developing multiple safeguards will be further studied.
	
%%%%%%%%%%%%%%%%%%%%%%%%%

	% or
	

	%% you can choose not to have a title for an appendix
	%% if you want by leaving the argument blank
	%\section{}

	%\section*{Acknowledgment}


	\ifCLASSOPTIONcaptionsoff
	\newpage
	\fi

	
	
	
	\bibliographystyle{IEEETran}
	\small
	%\bibliographystyle{IEEEtr}
	\bibliography{Reference}
	%\begin{thebibliography}{1}
	%\end{thebibliography}
	
	
	%\end{IEEEbiography}
	
	%\newpage
	
	
	
	%\vfill
	

	
\end{document}